\documentstyle[pra,aps]{revtex}
\input epsf
\draft

\begin{document}

\title{Two-Dimensional Sideband Raman Cooling and Zeeman State
Preparation in an Optical Lattice
\footnote{Contribution of NIST, not subject to copyright}}

\author{A.V. Taichenachev, A.M. Tumaikin, and V.I. Yudin}

\address{ Novosibirsk State
       University, Pirogova 2, Novosibirsk 630090, Russia}

\author{ L. Hollberg}

\address{ Time and Frequency Division,
       National Institute of Standards and Technology,\\
       325 Broadway, MS 847-10, Boulder, CO 80303}

\date{\today}

\maketitle

\begin{abstract}
A method of sideband Raman cooling to the vibrational
ground state of the $m=0$ Zeeman sublevel in a far-detuned
two-dimensional optical lattice is proposed. In our scheme, the Raman
coupling between vibrational manifolds of the adjacent Zeeman
sublevels is shifted to the red sideband due to the ac Stark effect
induced by a weak pump field. Thus, cooling and optical pumping to
$m=0$ is achieved by purely optical means with coplanar cw laser
beams. The optical lattice and cooling parameters
are estimated in the framework of simple theoretical models. An
application of the transverse sideband cooling method to frequency
standards is discussed. Coherent
population trapping for the sideband Raman transitions between the
degenerate vibrational levels is predicted.
\end{abstract}


\section{Introduction}

Laser-cooled atoms play a critical role in modern
frequency standards such as atomic fountains \cite{coldfs}.
As is well-known, Sisyphus-type cooling in optical molasses with
polarization gradients results in atoms with
temperatures corresponding to tens of the single-photon recoil
energies $\varepsilon_r = (\hbar k)^2/2M$ (for example,  $T \sim 30\,
\varepsilon_r/k_B \sim 3 \mu K$ in the case of $Cs$ \cite{salomon}).
Even lower temperatures can be achieved by velocity-selective
methods \cite{aspect88,chu92,lawall94}. These methods, however,
require more complicated technical implementations \cite{chu92}.

Recently, Poul Jessen and co-workers \cite{jessen98}
demonstrated an elegant and efficient method of cooling atoms to
the vibrational ground state of a far-off-resonance
two-dimensional optical lattice. Their method is a variant of
Raman sideband cooling \cite{sbcooling} based on transitions
between the vibrational manifolds of adjacent Zeeman substates.
A static magnetic field is used to tune the Zeeman levels so
that Raman resonance occurs on the red sideband and results in
cooling.  Two circularly polarized fields are then used to
recycle the atoms for repetitive Raman cooling.  The cooling
operates in the Lamb-Dicke regime with cw laser beams and does
not require phase-locked lasers; a transverse temperature of
about $950 \,nK$ was achieved.

Stimulated by the concepts and results from Jessen \cite{jessen98},
we propose a new variant of transverse sideband cooling. The basic
difference from the method of Ref. \cite{jessen98} is that a
linearly polarized pumping field, detuned from resonance, now plays a
two-fold role. It  provides both optical pumping back to the  $m = 0$
magnetic sublevel, and it causes a uniform ac Stark shift that
replaces the external magnetic field-induced Zeeman shift
that was used in Ref.
\cite{jessen98}.  The main improvement consists in the flexibility of
optical methods of the atomic state control. For example, in our
scheme, changing the polarization from linear to circular and the
direction of the pumping beam, we can accumulate all atoms in the
outermost Zeeman substate $m=F$.

One promising application of this technique is for
two-dimensional cooling of atoms in an atomic fountain, or for a
space-based atomic clock. More precisely, the method can be used
to cool an atomic sample down to sub-$\mu K$ temperatures in the
transverse directions before launching the cold atoms into the
Ramsey interaction region. For the long Ramsey periods that
could be used in an atomic clock in space, reducing the
transverse spread would increase substantially the number of
atoms in the detection region. In this context, Jessen's
original scheme has some disadvantages. Namely, atoms are
accumulated in the stretched $m = F$, substate of the $F=4$
ground-state hyperfine level of $Cs$, a perturbing magnetic
field is used, and that system also requires that the pumping
and repumping beams propagating orthogonal to the cooling plane
(i.e. down to axis of the clock), which would perturb atoms
drifting through the Ramsey region. Newly proposed
atomic-fountain clocks on Earth and clocks proposed for space
will use multiple balls of atoms in the Ramsey region, so there
would be atoms in the Ramsey cavity undergoing state
interrogation while other atoms are state prepared and detected.
In this case the out-of-plane light beams and magnetic field
would produce unacceptable frequency shifts.  Some of these
problems are of a technical character and could be solved by
different methods.  For instance, atoms might be transfered from
$|F=4, m=4\rangle$ to $|F=4, m=0\rangle$ without additional
heating by the adiabatic passage technique \cite{at}, and the
Ramsey region might be adequately isolated from the magnetic
field.  However, the problem of the light beams down the clock
axis is more fundamental problem.  Our scheme avoids these
problems without additional technical complications, while
maintaining most of the attractive features.  In particular, in
the present version, only cw lasers lying in the cooling plane
are used, there is no magnetic field, and atoms are prepared in
the $m=0$ substate. Since the linear polarization of the pumping
field coincides (must coincide) with the orientation of the
clock's magnetic field (C-field) in the Ramsey region, switching
between the Stark shifted and Zeeman shifted substates will not
produce any mixing.

In the following section we discuss the proposed lattice and
cooling parameters in the framework of a simple theoretical
model. The optimal magnitudes  of the Raman transition
amplitude, the pumping field intensity, and the detuning are
found. These analytical results are then confirmed by numerical
calculations for a more realistic model of the $F \rightarrow
F'=F$ transition. In addition, we find that coherence between
degenerate (or nearly degenerate) lower vibrational levels can
lead, under certain conditions, to significant changes in the
cooling efficiency and cooling time.

The proposed cooling
method may also be useful for atom optics  by providing
a high-brightness
well-collimated source of atoms, or for general purposes of
quantum-state control in a non-dissipative optical lattice.

\section{Optical Lattice}

The field configuration used for the optical lattice
consists of three linearly polarized beams having equal amplitudes and
propagating in the $xy$-plane with angles of $2 \pi/3$ between them
(Fig. 1). The polarization vectors of
these beams are tilted through a small angle $\phi$ with respect to
the $z$-axis.  The resulting field can be written as
\begin{eqnarray} \label{field}
{\bf E}({\bf r},t) &=& E_{0} {\bf {\cal E}}({\bf
r})\exp(-i\omega_L t) + {\rm c.c} \nonumber\\
{\bf {\cal E}}({\bf r}) &=&
{\bf e}_z \sum_{i=1}^{3} \exp(i {\bf k}_i{\bf r}) +
\tan(\phi)\sum_{i=1}^{3} {\bf e}_i\exp(i {\bf k}_i{\bf r}) \;,
\end{eqnarray}
where ${\bf k}_i$ and $\tan(\phi){\bf e}_i$ are respectively the wave
vectors and the in-plane components of the polarization of the $i$-th
beam.  All the beams have the same frequency, $\omega_L$,
far-detuned to the red of the $D_2$ resonance line.

As was shown in Ref. \cite{jessen97}, if the detuning is much greater
than the hyperfine splitting of the excited state, then the optical
potential for the ground state takes the form
\begin{equation} \label{pot_gen}
\widehat{U}_{F} = -\frac{2}{3} u_s |{\bf {\cal E}}({\bf{r}})|^2+
\frac{i}{3} u_s g(F)
[{\cal E}({\bf r})^* \times {\cal E}({\bf r})]
\cdot\widehat{{\bf F}} \;.
\end{equation}
Here $g(F)=[F(F+1)+J(J+1)-I(I+1)]/[F(F+1)]$ where
$F$, $J$ and $I$ are respectively the total, the electron and
the nuclear angular momenta of the ground state, and
$\widehat{{\bf F}}$ is the angular-momentum operator.
The single-beam light shift $u_s =- A
{\cal I}/\Delta$, defined as in Ref. \cite{jessen97}, is proportional to the
single-beam light intensity ${\cal I}$ and inversely proportional to
the detuning $\Delta = \omega_L -\omega_{F,F'_{max}}$. For the $D_2$ line of $^{133}Cs$ the constant $A
\approx 1.5\, \varepsilon_r \,GHz/(mW \cdot cm^{-2})$.

To zeroth order in $\tan(\phi) \ll 1$, the field (\ref{field})
is linearly polarized along ${\bf e}_z$ everywhere and vector term
in Eq. (\ref{pot_gen}) vanishes, resulting in the isotropic optical
potential
\begin{equation} \label{pot_zero}
\widehat{U}^{(0)} = -\frac{4}{3} u_s
\left[
\frac{3}{2}+ \cos(\sqrt{3} k x)+ \cos(\frac{\sqrt{3} k x- 3 k
y}{2})+ \cos(\frac{\sqrt{3} k x + 3 k y}{2}) \right] \;.
\end{equation}
In other words, contrary to the field configuration of Ref.
\cite{jessen98}, all the Zeeman sublevels have the same optical
shift. For red detunings $\Delta < 0$, the minima of the potential
(\ref{pot_zero})  form a lattice consisting of equilateral triangles
with a side $2 \lambda/3$ (one of them has the coordinates $x=y=0$).

In the general case, the atomic motion in a periodic potential leads
to a energy band-structure. However, for potentials with a
periodicity of the order of the light wavelength $\lambda$  and with
the depth  much larger than the recoil energy $\varepsilon_r$ ($ 6
u_s$ in the case under consideration), both the tunneling probability
and the width are exponentially small for bands close to a potential
minimum. Hence, instead of a lattice with energy bands we can consider
vibrational levels arising from independent potential wells.  The
spectrum of the lower levels can be defined, with good accuracy, from
the harmonic expansion in the vicinity of the well's bottom:
$$
\widehat{U}^{(0)} \approx u_s[-6+3 k^2 (X^2+Y^2)]\;,
$$
where $X$ and
$Y$ are the displacements from the minimum. This expansion
corresponds to a 2D isotropic harmonic oscillator with the frequency
$\hbar \omega_v = \sqrt{12 u_s \varepsilon_r}$. Due to the isotropy,
the $n$-th energy level is $n+1$ times degenerate. If the energy
separation between adjacent vibrational levels is much greater than
the recoil energy, the characteristic size of lower vibrational
states is $l = \sqrt{\hbar/M\omega_v} \ll \lambda$. In this case
we have strong localization, and the Lamb-Dicke regime holds.

\section{Sideband  Raman Cooling}

Raman transitions between vibrational levels of adjacent magnetic
substates are induced by the small in-plane component of the field
(\ref{field}). To first order in $\tan(\phi)$, the vector part
of Eqn. (\ref{pot_zero}) gives the correction
\begin{equation} \label{raman}
\widehat{U}^{(1)} = \frac{1}{3} u_s g(F) \tan(\phi)
{\bf M}({\bf r})\cdot\widehat{\bf F} \;,
\end{equation}
where ${\bf M}$ has the components $M_x=2\sqrt{3}[\cos(3 k
y/2)\sin(\sqrt{3} k x/2) + \sin(\sqrt{3} k x)]$ and $M_y=6 \sin(3 k
y/2)\cos(\sqrt{3} k x/2)$.  Since this term conserves the symmetry of
the main potential (\ref{pot_zero}), each well in the lattice obeys
the same conditions for the Raman transitions. For the
lower vibrational levels we use a first-order approximation with
respect to the displacements $X,\,Y$ from the minimum
\begin{eqnarray} \label{u1_harm}
\widehat{U}^{(1)}&\approx& 3 u_s g(F)\tan(\phi) k
(X\widehat{F}_x+Y\widehat{F}_y) \nonumber\\
&=&
\frac{3}{2} u_s g(F)\tan(\phi) k
\left((X-iY)\widehat{F}_{+}+(X+iY)\widehat{F}_{-}\right)
\;,
\end{eqnarray}
where $\widehat{F}_{\pm}$ are the standard raising and lowering
angular momentum operators.
The operator $\widehat{U}^{(1)}$ has off-diagonal elements both for
the vibrational and for the magnetic quantum numbers, inducing
transitions with the selection rules $\Delta n = \pm1$ and $\Delta m
= \pm 1$ (for a quantization axis along ${\bf e}_z$). In order of
magnitude, the Raman transition rate between the lower vibrational
levels is $U_R = u_s \tan(\phi) k l $.  As was shown in Refs.
\cite{jessen97,jessen98}, sideband cooling and coherent quantum-state
control require this rate to be much greater than the spontaneous
scattering rate of lattice photons $\gamma_s = 6 \Gamma u_s/\Delta$,
where $\Gamma$ is the natural width. In our lattice $U_R/\gamma_s
\approx 0.2 \tan(\phi) \Delta/\Gamma (\varepsilon_r/u_s)^{1/4}$.



Two other important requirements for efficient Raman sideband cooling
are a spatially independent energy shift of the magnetic sublevels
and optical pumping.  To achieve these, we propose to use another
optical field, known as the pump beam, linearly polarized along the
$z$-axis, propagating in the cooling plane, and detuned by several
$\Gamma$ to the blue of the $F \rightarrow F''=F$ transition of the
$D_1$ line \cite{reason} (Fig. 1).  In this case the $m = 0$
sublevel is dark and unshifted, while the others undergo the light
shifts $$ \delta_{m} = m^2 \frac{\Delta_p
\Omega_p^2}{\Gamma^2/4+\Delta_p^2} \;, $$ where $\Omega_p$ is the
Rabi coupling for the $|F,m=\pm 1\rangle \rightarrow |F''=F,m''=\pm
1\rangle$  transitions and $\Delta_p$ is the detuning of the pump
field. With a proper choice of $\Omega_p$ and $\Delta_p$, the states
$|m=0,n+1\rangle$ and $|m=\pm 1,n\rangle$ will have the same energy,
which leads to efficient transition between them due to the Raman
coupling. The cooling picture is completed by the optical pumping,
which provides the relaxation from $|m=\pm 1,n\rangle$ to
$|m=0,n\rangle$ (see Fig. 2.a). The vibrational quantum number $n$
is conserved in this process due to the fact that atoms are in the
Lamb-Dicke regime.  It is important to note
that, contrary to Ref.
\cite{jessen98}, in our case, several levels
are cooled simultaneously
due to the isotropy of the potential
$\widehat{U}^{(0)}$ (\ref{pot_zero}).  If $\omega_v \gg U_R$ the
state $|m=0, n=0\rangle$ is approximately dark and the majority of
the atoms are eventually pumped into this target state. Thus, the
described cooling method can be viewed as a version of dark-state
cooling. Aspects of this cooling scheme that distinguish it from that
of Ref. \cite{jessen98} are: the cooling will be more efficient
because more levels involved, the final pumping is to dark state
$m=0$, and only laser fields in a plane perpendicular to the
quantization axis are used.

To insure that there are no additional
constraints and to estimate the cooling parameters, we consider a
simple theoretical model based on the double $\Lambda$-system
(see Fig. 2.b). This simplified case
allows an analytical treatment of the problem. We find
the steady-state solution to the generalized optical Bloch
equations involving the light-induced and spontaneous transitions and
the Raman coupling.  We are primarily interested in the limits
\begin{equation} \label{limits}
\Delta_p \gg \Gamma\;;\;\; \omega_v \gg U_r  \;,
\end{equation}
because in this case the light-shift
exceeds the field broadening and we can shift the states $|2\rangle$
and $|5\rangle$ into degeneracy with negligible perturbation of state
$|1\rangle$.  Under these conditions, the solution
gives the following results:
(i) the population of the target state
$|1\rangle$ is maximal at exact resonance $\Omega_p^2/\Delta_p =
\omega_v$ (see Fig. 3.a),
(ii) on resonance, the total population of the states coupled with
light is small, and equal to $(U_R/\omega_v)^2$, the probability for
off-resonance Raman transitions $|1\rangle \rightarrow |6\rangle$
multiplied by a factor 4,
and (iii) the population of state $|2\rangle$ contains two
terms:  $\pi_2 =1/2 (U_R/\omega_v)^2+1/16 (\gamma_p/\omega_v)^2$.
The second term is determined by the ratio of the width imposed by
light, $\gamma_p = \Gamma \Omega_p^2/\Delta_p^2$, to the vibrational
frequency $\omega_v$.  As a result, the target state population is
close to unity:
\begin{equation} \label{target}
\pi_1 \approx 1 - a(U_R/\omega_v)^2 - b (\Gamma/\Delta_p)^2 \;.
\end{equation}
The coefficients are $a=3/2$ and $b=1/16$ in the case of the
double-$\Lambda$ system model.

We now turn to an estimate of the cooling dynamics. Instead of
looking for a temporal solution of the Bloch equations
$(d/dt)\rho = \widehat{{\cal L}}\rho$
for atomic density matrix $\rho$
and $\widehat{{\cal L}}$ the corresponding Liouvillian superoperator,
we find the
statistically averaged transition time $\tau =
\int_{0}^{\infty}(\rho(t)-\rho(\infty))dt$ \cite{APW}.  This matrix
obeys the equations
$\widehat{{\cal L}}\tau = \rho(\infty)-\rho(0)$, where
$\rho(\infty)$ is the steady-state solution and $\rho(0)$ is the
initial distribution (we set $\pi_1=\pi_2=\pi_5=\pi_6=1/4$ and the
other elements equal to zero at $t=0$).  The cooling rate can be
associated with the inverse transition time for the $|1\rangle$ state
$\gamma_{cool} = \tau_1^{-1}$.  As a function of the optical
frequency shift, the cooling rate is a Lorentzian curve with a
width $\sim \sqrt{1/4 \gamma_p^2+ 7 U_R^2}$ (see Fig. 3.a). Exactly
on resonance,  and in the limits
(\ref{limits}) $\Delta_p = \Omega_p^2/\omega_v$,
and $\gamma_{cool}$ takes the form
\begin{equation} \label{rate}
\gamma_{cool} = \alpha \frac{\gamma_p U_R^2}{\gamma_p^2+\beta U_R^2} \;.
\end{equation}
Calculations within
the framework of the double $\Lambda$-system give $\alpha = 8$ and
$\beta = 28$. This dependence of the cooling rate on
$\gamma_p$ and $U_R$ can be
understood qualitatively if we consider
the cooling as optical pumping
into the dark state. Obviously, because
other parameters do not appear in the conditions (\ref{limits}),
the cooling rate is determined entirely by the optical pumping rate
$\gamma_p$ and the Raman transitions rate $U_R$.
If $U_R \gg \gamma_{p}$, an atom
passes from $|2\rangle$ to $|5\rangle$ very quickly, and the cooling
rate is proportional to the rate of the slower process of repumping
from $|5\rangle$ to $|1\rangle$. In the inverse limit $U_R \ll
\gamma_{p}$, the slowest process is the transition
$|2\rangle \rightarrow |5\rangle$.
The corresponding rate, however, is not equal to $U_R$, but is
suppressed by the factor $U_R/\gamma_{p}$. That can be explained as
the inhibition of quantum transitions due to continuous measurements
on the final state $|5\rangle$ (quantum Zeno effect \cite{zeno}).
The cooling rate $\gamma_{cool}$ as a function of $\Omega_p$ (on
resonance) is shown in Fig.3.b; $\gamma_{cool}$ achieves a maximum
$\gamma_{cool}^{max} = U_R \alpha/(2\sqrt{\beta})$ at the optimal
Rabi coupling
$\Omega_p^{opt} = (\beta)^{-1/4} \omega_v \sqrt{\Gamma/U_R}$.

The above described laws for the target-state population and for the
cooling dynamics are confirmed by numerical calculations for a more
realistic model of the $F \rightarrow F'=F$ cycling transition with a
limited number of vibrational levels of the 2D oscillator taken into
account. More precisely, we consider the systems of equations for the
steady-state density matrix $\rho(\infty)$ and for the transition
time matrix $\tau$ taking into account for all possible coherences and
populations.  Including all terms gives a large
system of algebraic equations, which
can be solved for a fixed set of parameters by numerical methods. We
use the LU decomposition method. The numerical results are fitted by
the formulae (\ref{target},\ref{rate}) very well.  The fitting
coefficients $a$, $b$, $\alpha$ and $\beta$ depend on the angular
momentum $F$ and on the initial distribution among the vibrational
levels.  The results, corresponding to the three lowest vibrational
levels (with initially equal populations), are presented in Table 1.

In principle, two factors limit the number of
vibrational levels which participate efficiently in the cooling:
both the anharmonicity and the violation of the
Lamb-Dicke regime become appreciable for higher vibrational levels.
The second factor is the more stringent limitation and gives
$n^* \approx 0.1\, \hbar \omega_v/\varepsilon_r$ as an
estimate for the maximal vibrational number.

It should be noted that a stray magnetic field can limit the cooling
efficiency by breaking the level degeneracy.
To be negligible, the Zeeman shift due to any stray magnetic field
should be less than the Raman transition rate $U_R$, or the width
imposed by the pumping field $\gamma_p$, which are of the same order
of magnitude for optimal cooling.

To provide the cyclic interaction of the atoms with the
pump field, repumping from the other hyperfine level is
necessary. For this
we propose using another light beam tuned in resonance
with a $F \rightarrow F'=F+1$ transition of the $D_2$ line.  This
beam is linearly  polarized along $e_z$  and runs in the $xy$-plane.
For example, if the pumping field operates on the $F=4 \rightarrow
F''=4$ of the $D_1$ line of $Cs$, the repumping field is applied to
the $F=3 \rightarrow F'=4$ transition of the $D_2$ line.  To minimize
effects of optical pumping on the other hyperfine level, the
intensity of the repumping field should be chosen close to the
saturation intensity. It is noteworthy that in our lattice the
potentials for both hyperfine levels have the same spatial
dependence, and consequently the requirement on the repump
intensity is not as stringent as in Ref.  \cite{jessen98}.


For purposes of illustration we can
give numerical estimations for $^{133}Cs$ ($\Gamma \approx
2\pi\, 5\, MHz$ and $\varepsilon_r/\hbar \approx 2 \pi \, 2\, kHz$).
If we take the lattice beams detuning, $\Delta = - 2\pi\,10 \,GHz$
(from the $F=4 \rightarrow F'=5$ transition of the $D_2$ line) and
intensity ${\cal I} = 500 \,mW/cm^2$, then the single-beam Stark
shift $u_s \approx  75\, \varepsilon_r \approx 2 \pi\hbar\, 150 \, kHz$.
The lattice has the depth $6 u_s = 450\, \varepsilon_r \approx 2\pi\hbar\,
900 \,kHz$, and thus supports
approximately $15$ bound levels with the energy
separation $\hbar \omega_v = 30 \, \varepsilon_r \approx 2 \pi\hbar\, 60
\, kHz$.  With the tilt angle $\tan(\phi) \approx 0.1$, the Raman
transition rate is $U_R \approx 0.1\, \hbar \omega_v$, providing the
figure of merit $U_R/\gamma_s \approx 12 \gg 1$. The stray magnetic
field must be controlled to the level of a few $mG$ or less.
In the case of $Cs$ the
pumping field should
be applied to the $F=4 \rightarrow F''=4$ transition of
the $D_1$ line.
The repumping field should be tuned to
resonance on the $F=3 \rightarrow F'=4$
transition of the $D_2$ line,
and have an intensity $\sim 10 \, mW/cm^2$
in order to saturate the transitions from all Zeeman sublevels.
The optimal pumping field
detuning $\Delta_p \approx 0.2\,\Gamma \omega_v/U_R \approx 2\,
\Gamma$ and intensity ${\cal I}_p \approx 8\, mW/cm^2$ give the
cooling rate $\gamma_{cool} \approx 0.4\, U_R \approx 2 \pi\, 2.2 \,
kHz$, and time of $\tau \approx \gamma^{-1}_{cool}
\approx 10^{-4} s$. For these conditions our analysis predict
approximately $95 \%$ of the population will accumulate
in the target state $|F=4, m=0, n=0\rangle$ with lower
levels having vibrational numbers up to $n^* \approx 0.1\, \hbar
\omega_v/\varepsilon_r \approx 3$.

\section{Coherence between Vibrational Levels}

In the case of a symmetric field
configuration for 2D and 3D lattices a degeneracy of the vibrational
energy level structure occurs. For a 2D lattice (for example the
field configuration of Ref. \cite{jessen98} and our configuration)
in the harmonic approximation, the $n$-th vibrational
level contains $n+1$ sublevels $\{ |m,n_x+n_y=n\rangle\}$.
We find that the coherence induced between the
degenerate or near-degenerate vibrational levels can play an
important role, and
significantly change the efficiency of the
resonance Raman
coupling. For instance, consider a field configuration that differs
from the scheme presented in Fig. 1.a by the direction and
polarization of the pumping beam. Let this circularly polarized beam
propagate  along the $z$-axis as is shown in Fig. 4.a. In this
case atoms are optically pumped to the outermost Zeeman substate
$|m=F \rangle$ (see in Fig. 4.b). If we consider two degenerate
vibrational levels, for example $|m=F, n_x=1,n_y=0\rangle$ and
$|m=F, n_x=0, n_y=1\rangle$ coupled by Raman transitions with the
unique state $|m=F-1,n_x=n_y=0\rangle$ (Fig. 5.a),
we find
that there exists a superposition of the states
of the degenrate vibrational level
$$|\Psi_{NC}\rangle=|m=F, n_x=0,n_y=1\rangle+i|m=F,
n_x=1,n_y=0\rangle$$
that is not coupled with
$|m=F-1,n_x=n_y=0\rangle$ by the operator $\widehat{U}^{(1)}$
(\ref{u1_harm}):
$\widehat{U}^{(1)} |\Psi_{NC}\rangle = 0$.
Hence, part of the population will be trapped
in this superposition state, in an analogy with well-known coherent
population trapping in the $\Lambda$-scheme \cite{CPT}.
The result of such trapping is that the cooling efficiency is reduced,
i.e. the part of atoms accumulated in the target state
$|m=F,n_x=n_y=0\rangle$ will be significantly less than unity.
In the case of coupling between higher levels $|m=F,
n_x+n_y=n\rangle$ and $|m=F-1, n_x+n_y=n-1\rangle$, there always
exists a coherent superposition of the sublevels $|m=F,
n_x+n_y=n\rangle$, for which the operator of the Raman transitions is
equal to zero, as it is for the light-induced $\Lambda$-chains
\cite{yudin89}. However, for higher
vibrational levels the anharmonicity has to be taken into account and
the degeneracy is partially broken.
If the vibrational energy levels are only nearly degenerate,
due to either the anharmonicity or the optical lattice asymmetry,
dark superpositions are not completely stable due to
small energy splitting and  atoms  pass eventually  from
$|m=F, n_x+n_y=n\rangle$ to
$|m=F-1, n_x+n_y=n-1\rangle$. But the corresponding transfer
time will be greater than
the normal $U_R^{-1}$, leading to an increase of cooling time.

In the scheme presented in Fig. 1.a, this unwanted coherence effect
is avoided by the simultaneous Raman coupling of the two
states $|m=0,n_x+n_y=1\rangle$ of the degenerate vibrational level,
with the two other states $|m=\pm 1,
n_x=n_y=0\rangle$ with different amplitudes, as is shown in Fig.
5.b. It can be seen from this figure that the dark superposition for
the transition with $\Delta m =-1$ is
$$
(X+iY)\widehat{F}_{-}|\Psi_{NC}^{(-)}\rangle = 0 \,,\;\;\;
|\Psi_{NC}^{(-)}\rangle=|m=0, n_x=0,n_y=1\rangle+
i|m=0,n_x=1,n_y=0\rangle\,.
$$
While for the transition with $\Delta m =+1$ the uncoupled state is
given by
$$
(X-iY)\widehat{F}_{+}|\Psi_{NC}^{(+)}\rangle = 0 \,,\;\;\;
|\Psi_{NC}^{(+)}\rangle=|m=0, n_x=0,n_y=1\rangle-
i|m=0,n_x=1,n_y=0\rangle\,.
$$
The states $|\Psi_{NC}^{(-)}\rangle$ and
$|\Psi_{NC}^{(+)}\rangle$ are orthogonal, so no a superposition
can be made
which nullifies the Raman coupling operator
$\widehat{U}^{(1)}$ (\ref{u1_harm}).
We also note that the coherence
within the vibrational structure might be very useful for other
purposes, for instance in quantum state preparation.

\section{Conclusion}

Concluding, we have proposed a new scheme for 2D
Raman-sideband-cooling to the zero-point energy in a
far-off-resonance optical lattice and simultaneously pumping atoms
to the $m=0$ dark state. The main distinguishing features of our
proposal is that the method uses a pumping field to
Stark-shift the
Raman coupling to the red sideband and thus accumulation cold
atoms in the  $m=0$ Zeeman sublevel. An elementary theoretical
consideration allowed us to calculate the dependence
for the cooling efficiency and for the cooling dynamics.
Our estimates for a cold $Cs$ show
that for easily
realizable experimental parameters
up to $95\%$ of atoms can be accumulated in the $|F=4, m=0,
n=0\rangle$ state in a millisecond time scale. This corresponds to
a kinetic temperature on the order of $100\, nK$ after adiabatic
release from the lattice \cite{adcooling}.  Besides,
coherent population trapping
between degenerate vibrational levels has been predicted
for the sideband Raman transitions.

The authors thank Dr. J. Kitching and Prof. P. Jessen for helpful
discussions.
This work was supported in part by the Russian Fund for Basic Research
(Grant No. 98-02-17794).
AVT and VIYu acknowledge the hospitality of NIST, Boulder.


\begin{table}
\caption{The fitting parameters for different transitions $F
\rightarrow F''=F$.}
\begin{center}
\begin{tabular}{ccccc} \hline \hline  \\
$F$ & $a$ & $b$ & $\alpha$ & $\beta$\\
\hline\hline
1 & 3.0 & 0.13 & 6.0 & 27 \\
2 & 2.6 & 0.13 & 4.6 & 27 \\
3 & 2.5 & 0.13 & 4.1 & 27 \\
4 & 2.4 & 0.13 & 3.8 & 27 \\
\hline
\hline
\end{tabular}
\end{center}
\end{table}

\begin{figure}
\epsfxsize=15 cm
\epsfbox{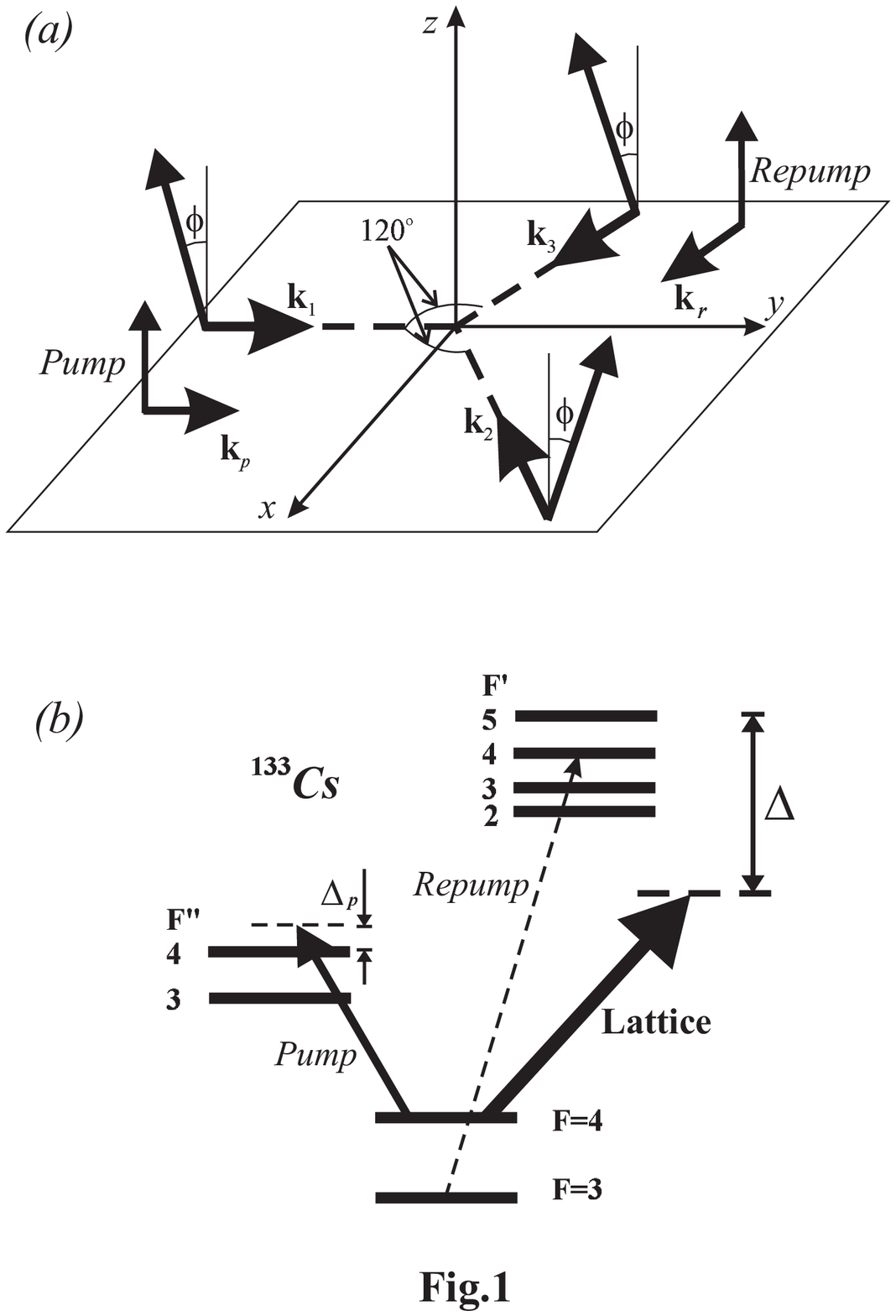}
\caption{(a) Field geometry. The basic optical lattice is formed by
three coplanar beams that are
linearly polarized along the $z$ -axis. The small
in-plane component of polarizations (the tilt angle is indicated by
$\phi$) induces the Raman coupling.
The pumping and repumping beams also propagate
in the $xy$-plane and
have $e_z$ linear polarization.
(b) The field detunings for the case of $^{133}Cs$ are: the lattice
beams are far-detuned to the red side of the $F=4 \rightarrow
F'_{max}=5$ transition of the $D_2$ line, the pumping field is tuned
to the blue side of the $F=4 \rightarrow F''=4$ transition of the
$D_1$ line, and the repumping  field is tuned in resonance with the
$F=3 \rightarrow F'=4$ transition of the $D_2$ line.}
\end{figure}

\begin{figure}
\epsfxsize=15 cm
\epsfbox{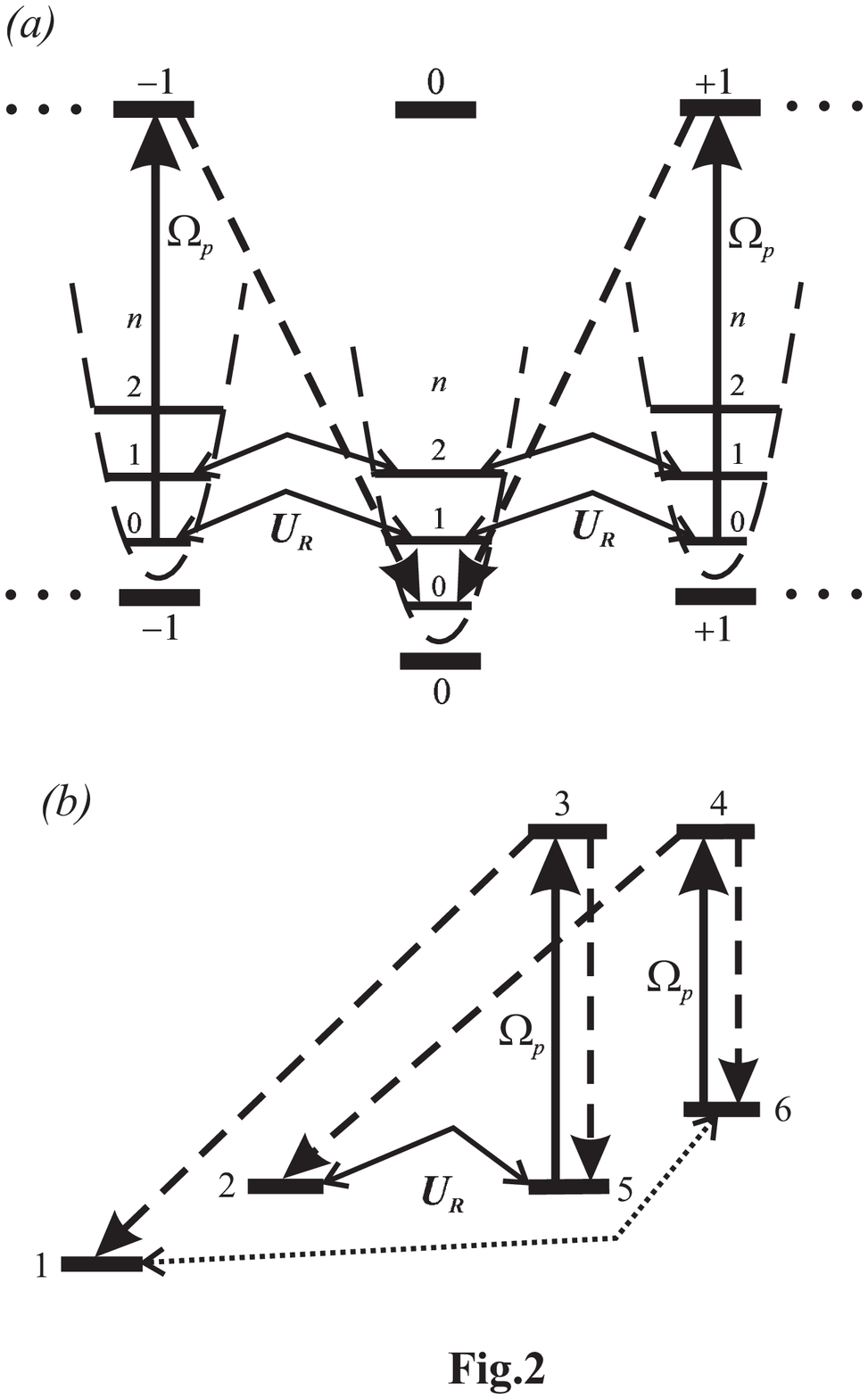}
\caption{(a) Scheme of the sideband Raman cooling.
Raman
transitions are shown by tilted arrows, connecting $|m=0, n\rangle$
to $|m=\pm 1, n'=n-1\rangle$ states.
The transitions induced
by the pumping field (solid lines) and spontaneous transitions (dashed
lines) provide the relaxation back to the $|m=0, n=0\rangle$
state.  The Zeeman sublevels are shown shifted by the optial Stark
effect of the pump laser field.
(b) Simple double-$\Lambda$ system model. The states $|1\rangle$,
$|3\rangle$, and $|5\rangle$ are within the odd
$\Lambda$-system and are connected by the radiative transitions,
which are shown by solid lines.  Similarly, the radiative transitions
within the even set $|2\rangle$, $|4\rangle$, $|6\rangle$
are shown by dashed lines.
Transitions between the
$\Lambda$-systems $|2\rangle \leftrightarrow |5\rangle$  and
$|1\rangle \leftrightarrow |6\rangle$ are both caused by the Raman
coupling.  (The dots indicate the off-resonant transitions $|1\rangle
\leftrightarrow |6\rangle$)}
\end{figure}

\begin{figure}
\epsfxsize=15 cm
\epsfbox{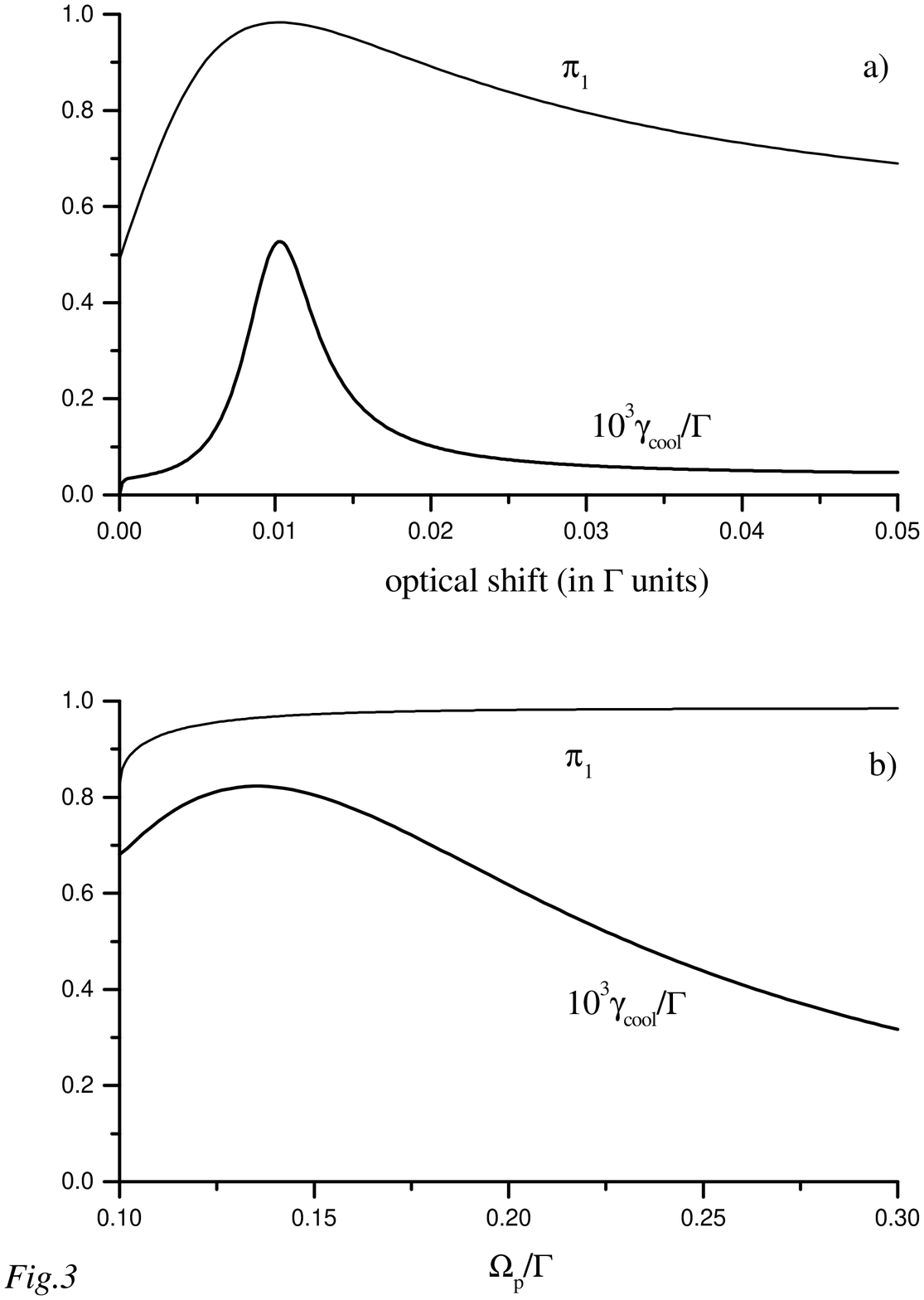}
\caption{ (a) The target state population $\pi_1$ and the
cooling rate are plotted versus
 the optical shift in the case of the double-$\Lambda$ system model
for parameters $\Delta_p = 5 \,\Gamma$,
$\omega_v = 0.01\, \Gamma$ and $U_R = 0.001 \, \Gamma$.  (b) The
target state population $\pi_1$ and the cooling rate
are plotted versus the Rabi
frequency for exact resonance. The parameters are $\omega_v = 0.01\,
\Gamma$ and $U_R = 0.001 \, \Gamma$.}
\end{figure}

\begin{figure}
\epsfxsize=15 cm
\epsfbox{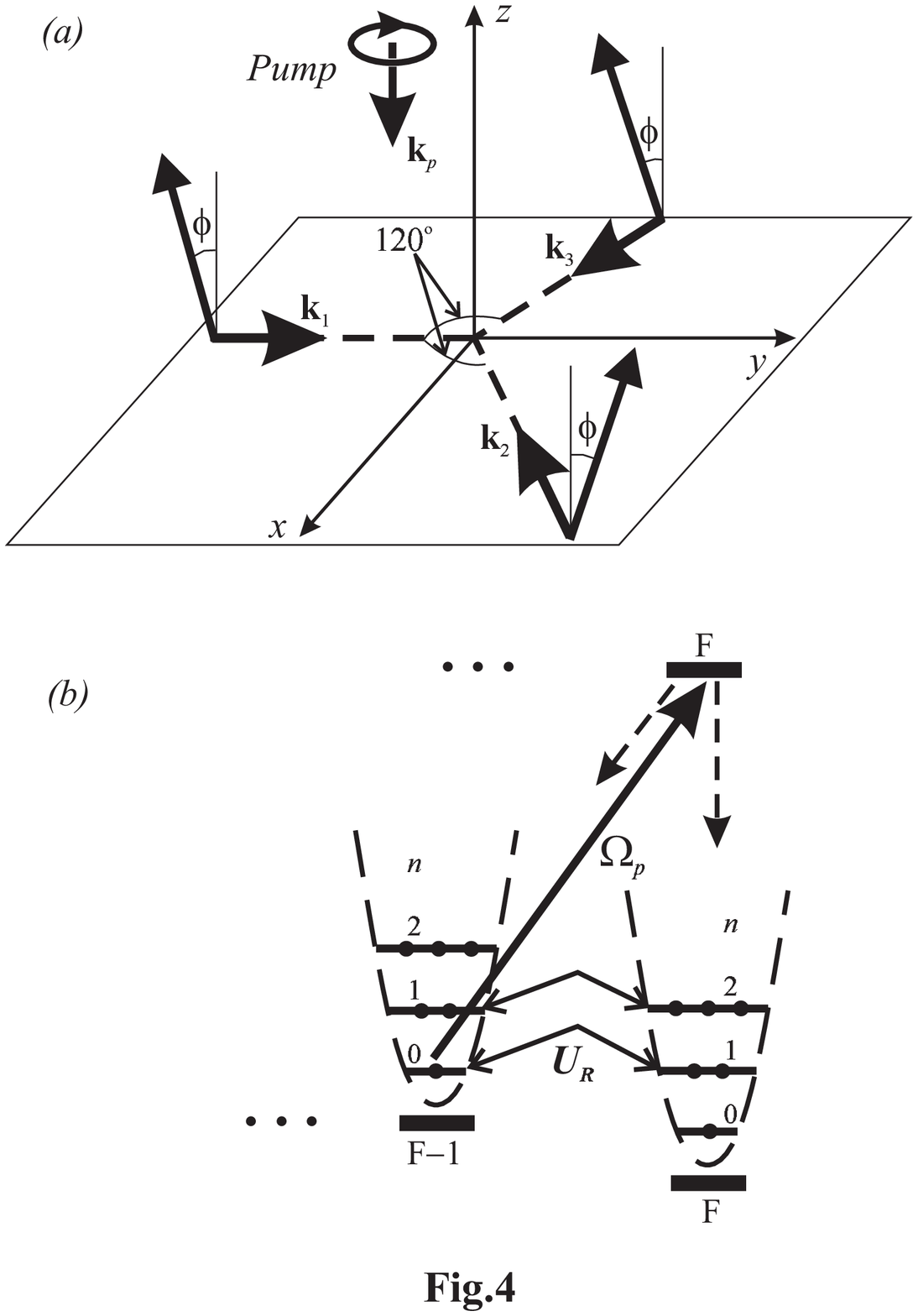}
\caption{(a) Field configration for the $m=F$ Zeeman state
preparation. It differs from the scheme shown
in Fig. 1.a by the
polarization and direction of the pumping beam.
(b) The light-induced, spontaneous, and Raman transitions between the
vibrational levels of Zeeman substates
for this case.
The small circles on the vibrational levels indicate the
degree of degeneracy.}
\end{figure}

\begin{figure}
\epsfxsize=15 cm
\epsfbox{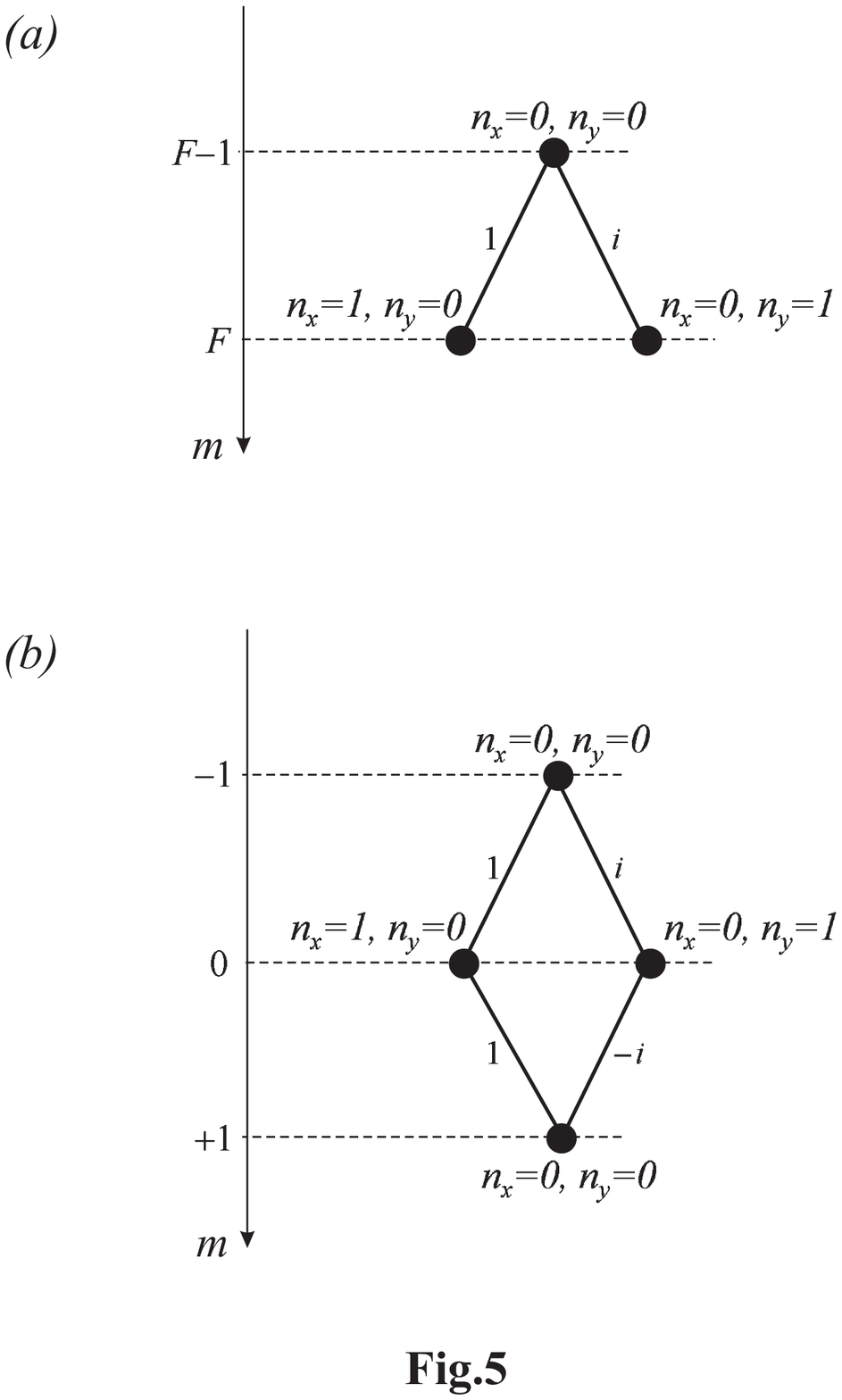}
\caption{Scheme of Raman coupling between the lower vibrational
levels of adjacent Zeeman substates  taking into account
the degeneracy
(a) in the case of the $m=F$  state preparation, and (b) in the case
of the $m=0$  state preparation. The relative amplitudes of the Raman
transitions (solid lines) are indicated.}
\end{figure}


\begin{thebibliography}{14}
\bibitem{coldfs}
J. J. Bollinger, J. D. Prestage, W. M. Itano, and D.  J. Wineland,
Phys. Rev. Lett., {\bf 54}, 1000 (1985);
M. A. Kasevich, E. Riis, S. Chu, and R. G. DeVoe, Phys. Rev. Lett.,
{\bf 63}, 612 (1989);
K. Gibble and S. Chu, Phys. Rev. Lett., {\bf 70}, 1771 (1993).
\bibitem{salomon} C. Salomon, J. Dalibard, W. D. Phillips, A.
Clairon, and S. Guellati, Europhys. Lett., {\bf 12}, 683 (1990).
\bibitem{aspect88}
A. Aspect, E. Arimondo, R. Kaiser, N. Vansteenkiste, and
C. Cohen-Tannoudji, Phys.  Rev.  Lett., {\bf 61} (1988) 826.
\bibitem{chu92} M. Kasevich and S. Chu, Phys. Rev. Lett.,{\bf 69},
1741 (1992).
\bibitem{lawall94} J. Lawall, F. Bardou, B. Saubamea, K. Shimizu, M.
Leduc, A. Aspect, and C. Cohen-Tannoudji, Phys. Rev. Lett., {\bf 73},
1915 (1994).
\bibitem{jessen98} S. E. Hamann, D. L. Haycock, G. Klose, P. H. Pax,
I. H. Deutsch, and P. S. Jessen, Phys. Rev. Lett., {\bf 80}, 4149
(1998).
\bibitem{sbcooling}
D. J. Heinzen and D. J. Wineland, Phys. Rev. A, {\bf 42}, 2977
(1990);
R. Ta\"{\i}eb, R. Dum, J. I. Cirac, P. Marte, and P. Zoller,
Phys. Rev. A, {\bf 49}, 4876 (1994);
H. Perrin, A. Kuhn, I. Bouchoule, and C. Salomon, Europhys. Lett.,
{\bf 42}, 395 (1998).
\bibitem{at}
P. Pillet, C. Valentine, R.-L. Yuan, and J. Yu, Phys. Rev. A,
{\bf 48}, 845 (1993).
\bibitem{jessen97} I. H. Deutsch and P. S. Jessen, Phys. Rev. A, {\bf
57}, 1972 (1997).
\bibitem{reason} We propose to use the $D_1$ line in order to avoid
any interference with the repumping and lattice beams, which operate
on the $D_2$ line.
\bibitem{APW} This method is a variant of a
statistical consideration of a dynamical system first introduced in
L. S. Pontryagin, A. A. Andronov, and A. A. Witt, Zh. Eksp. Teor.
Fiz., {\bf 3}, 165 (1933).
\bibitem{zeno} W. M. Itano, D. J. Heinzen, J. J. Bollinger, and D. J.
Wineland, Phys. Rev. A, {\bf 41}, 2295 (1990).
\bibitem{CPT} E. Arimondo, in {\em Progress in Optics}, edited
by E. Wolf (North-Holland, Amsterdam, 1996), V. XXXV, p.259.
\bibitem{yudin89}
V. S. Smirnov, A. M. Tumaikin, and V. I. Yudin,
Sov. Phys. JETP,
{\bf 69}, 913, (1989).
\bibitem{adcooling} A. Kastberg, W. D. Phillips, S. L. Rolston, R. J.
C. Spreeuw, and P. Jessen, Phys. Rev. Lett., {\bf 74}, 1542 (1995).

\end{thebibliography}
\end{document}